\documentclass[prl,twocolumn,fleqn]{revtex4}
\usepackage{epsfig}
\usepackage{graphicx}

\usepackage{amsmath}
\usepackage[psamsfonts]{amssymb}
\usepackage{euscript}

\usepackage{latexsym}

\def\hsp{,\hspace{.7cm}}

\renewcommand{\(}{\begin{equation}}
\renewcommand{\)}{end{equation} \vspace{-.05in}\linebreak}

\newcounter{saveeqn}
\newcounter{savealpheqn}

\newcommand{\alpheqn}{\setcounter{saveeqn}{\value{equation}}%
  \stepcounter{saveeqn}\setcounter{equation}{0}%
  \renewcommand{\theequation}{\mbox{\arabic{section}.\arabic{saveeqn}
\alph{equation}}}
  \renewcommand{\)}{\end{equation}}}
\def\part#1{\frac{\partial}{\partial{#1}}}%
\def\group#1{\refstepcounter{equation}\setcounter{saveeqn}
 {\value{equation}}%
  \label{#1}\setcounter{equation}{0}%
\renewcommand{\theequation}{\mbox{\arabic{section}.\arabic{saveeqn}
\alph{equation}}}
  \renewcommand{\)}{\end{equation}}}
\newcommand{\reseteqn}{\setcounter{equation}{\value{saveeqn}}%
  \renewcommand{\theequation}{\arabic{section}.\arabic{equation}}%
  \renewcommand{\)}{\end{equation}}}

\newcommand{\aalpheqn}{\setcounter{saveeqn}{\value{equation}}%
  \stepcounter{saveeqn}\setcounter{equation}{0}%
  \renewcommand{\theequation}{\mbox{
        \Alph{subsection}.\arabic{saveeqn}\alph{equation}}}
   \renewcommand{\)}{\end{equation}}}
\newcommand{\areseteqn}{\setcounter{equation}{\value{saveeqn}}%
  \renewcommand{\theequation}{\Alph{subsection}.\arabic{equation}}%
  \renewcommand{\)}{\end{equation}}}

\renewcommand{\thefootnote}{\alph{footnote}}
\renewcommand{\(}{\begin{equation}}
\renewcommand{\)}{\end{equation}}
\newcommand{\ba}{\begin{eqnarray}}
\newcommand{\ea}{\end{eqnarray}}

\newcommand{\bp}{\mathop{\vtop{\ialign{##\crcr
   $\hfil\displaystyle{}\hfil$\crcr\noalign{\kern-13pt\nointerlineskip}
   \BIG{(}\hskip0pt\crcr\noalign{\kern3pt}}}}}
\newcommand{\cbp}{\mathop{\vtop{\ialign{##\crcr
   $\hfil\displaystyle{}\hfil$\crcr\noalign{\kern-13pt\nointerlineskip}
   \BIG{)}\hskip0pt\crcr\noalign{\kern3pt}}}}}
\newcommand{\pa}{\mathop{\vtop{\ialign{##\crcr
    
$\hfil\displaystyle{\oplus}\hfil$\crcr\noalign{\kern+1pt\nointerlineskip 
}
   \hspace{.08in}$^{\alpha=0}$\hskip6pt\crcr\noalign{\kern3pt}}}}}
\renewcommand{\hsp}{,\hspace{.3in}}

\newcommand{\beq}{\begin{equation}}
\newcommand{\eeq}{\end{equation}}

\numberwithin{equation}{section}
\renewcommand{\theequation}{\mbox{\arabic{equation}}}

\def\hsp#1{\hspace{#1in}}

\catcode`\@=11
\def\vereq#1#2{\lower3pt\vbox{\baselineskip1.5pt \lineskip1.5pt
\ialign{$\m@th#1\hfill##\hfil$\crcr#2\crcr\sim\crcr}}}
\catcode`\@=12

\makeatletter
\newcommand\figcaption{\def\@captype{figure}\caption}
\newcommand\tabcaption{\def\@captype{table}\caption}
\makeatother
\renewcommand{\(}{\begin{equation}}
\renewcommand{\)}{\end{equation}}

\renewcommand{\beq}{\begin{equation}}
\renewcommand{\eeq}{\end{equation}}
\newcommand{\bea}{\begin{eqnarray}}
\newcommand{\eea}{\end{eqnarray}}
\newcommand{\beas}{\begin{eqnarray*}}
\newcommand{\eeas}{\end{eqnarray*}}

\newcommand{\bquo}{\begin{quote}}
\newcommand{\enqu}{\end{quote}}
\def\hsp{,\hspace{.2cm}}

\begin{document}
% ======================================================================== 
\def\thefootnote{\fnsymbol{footnote}}

\title{What can Gaia (with TMT) say about Sculptor's Core?}

\author{Jarah Evslin}
\affiliation{Institute of Modern Physics, CAS, NanChangLu 509, Lanzhou 730000, China
}

%\author{Manoj Kaplinghat (coauthor?)}
%\affiliation{Department of Physics and Astronomy, UC Irvine, CA 92697, USA}

\begin{abstract}
\noindent
Walker et al.'s Magellan/MMFS Survey survey identified 1355 red giant candidates in the dwarf spheroidal galaxy Sculptor.  We find that the Gaia satellite will be able to measure the proper motions of 139 of these with a precision of between 13 and 20 km/s.  Using a Jeans analysis and 5-parameter density model we show that this allows a determination of the mass within the deprojected half-light radius to within 16\% and a measurement of the dark matter density exponent $\gamma$ to within 0.68 within that radius. If, even at first light, the TMT observes Sculptor then the combined observations will improve the precision on these proper motions to about 5 km/s, about 5 years earlier than would be possible without Gaia, further improving the precision of $\gamma$ to 0.27.  Using a bimodal stellar population model for Sculptor the precision of $\gamma$ improves by about 30\%.  This suggests that Gaia (with TMT) is capable of excluding a cusped profile of the kind predicted by CDM simulations with 2$\sigma$ (4$\sigma$) of confidence.

\end{abstract}

% \vfill
% 
% \end{titlepage}
\setcounter{footnote}{0}
\renewcommand{\thefootnote}{\arabic{footnote}}

% \pacs{??}

\maketitle

According to the standard cosmological model, most of the matter in the universe is cold dark matter (CDM), which at long distances only interacts gravitationally.  Direct and indirect dark matter detection experiments are a major industry and so far have rapidly excluded large regions of the CDM parameter space, as has the Large Hadron Collider.  However such tests can only confirm CDM, any such attempt to falsify this paradigm can be evaded by fine-tuning.  On the other hand, CDM makes falsifiable predictions for the density profiles of {\it{sufficiently}} dark matter dominated systems.  They must fall as $1/r^3$ at large radii and as at least $1/r$ at small radii, a scaling known as a cusp.  While there is little consensus over just how much dark matter domination is sufficient, studies such as Ref.~\cite{governato,gnatura} suggest that galaxies with stellar masses beneath about $10^{6.5-7}\ M_\odot$ should be cusped while Ref.~\cite{pencons} argues that energy conservation implies that baryonic effects cannot remove cusps in the Milky Way's dwarf spheroidal (dSph) satellites except under quite special circumstances.  Thus the discovery that a dSph is cored, not cusped, would pose a strong challenge to CDM.

In this letter, following a strategy similar to that of Ref.~\cite{strigarifisher}, we determine just when it will be known whether the dSph Sculptor, with a stellar mass of $2.3\times 10^6\ M_\odot$, is cored.  Common wisdom states that one need first wait 10 years for a 30 meter class telescope, such as the Thirty Meter Telescope (TMT), to observe such a galaxy.  Then one measures the positions of the stars, waits 5 {\it{more}} years and observes again.  This yields proper motions of hundreds or thousands stars at a precision of about 4 km/sec, allowing a clean resolution to the cusp/core problem around 2030.  We will argue that the Gaia satellite can yield a 2$\sigma$ hint of the cored/cusped nature of Sculptor within 5 years and then, when its position measurements are combined with TMT's {\it{first}} observations of Sculptor in under 10 years, a definitive exclusion of a cusped profile will already be possible.

\begin{figure} %[!tph]
\begin{center}
\includegraphics[width=2.8in,height=1.2in]{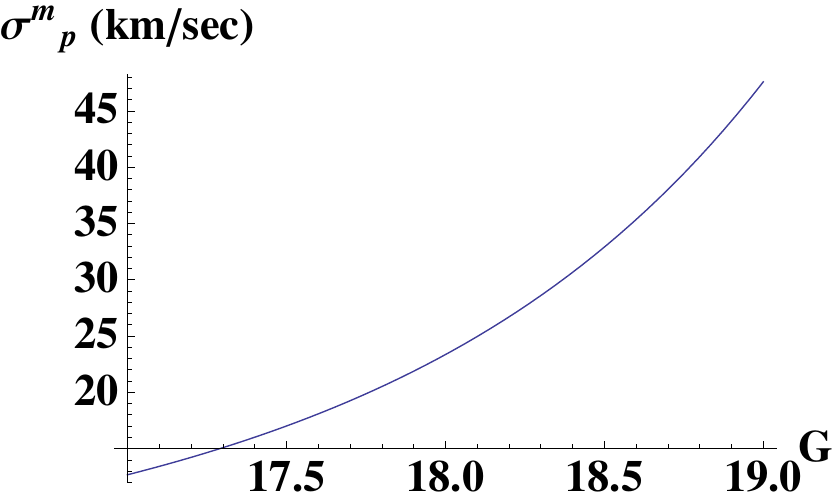}
\caption{The precision with which Gaia can measure the proper motion of a Sculptor member with Gaia magnitude $G$ and color $V-I_C=1.35$}
\label{sigfig}
\end{center}
\end{figure}

In Ref.~\cite{mmfs} the authors report the observation of 1541 objects in the part of the sky occupied by the dSph Sculptor, as part of their MMFS survey using the Magellan/Clay Telescope \footnote{This survey includes 3 other dSphs, but due to the distance to Fornax and the lack of bright stars in the others, Sculptor allows the best determination of the dark matter halo profile.}.  Of these, we consider the 1355 objects for which they assign a 90\% or greater membership probability in Sculptor.   For each object they provide the $V$ magnitude and the $I$ magnitude, which we identify with the Johnson-Cousins magnitude $I_C$.  We use these to calculate the Gaia magnitude, $G$, following Ref.~\cite{gmag}
\bea
G&=&V-0.0257-0.0924(V-I)\nonumber\\&&-0.1623(V-I)^2+0.0090(V-I)^3 .
\eea
As described in the Gaia Science Performance document \cite{gaiaperf} and updated in \cite{gaia2015} to reflect post-launch performance, Gaia can measure proper motions with an end of mission precision, in $\mu$as/yr, of 
\bea
\sigma^{m0}_p\hspace{-.2cm}&=&\hspace{-.2cm}\sqrt{-1.6+680.8 (10)^{0.4(G-15)}+32.7 (10)^{0.8(G-15)}}\nonumber\\&&\times 0.526\left(0.986+(1-0.986)(V-I)\right). \label{prec}
\eea
There are also position-dependent corrections to this precision.  Using the location of Sculptor and the precision corrections expected~\cite{gaiaperf}, we find an improvement in this precision of roughly 7 $\mu$as/yr (14 $\mu$as/yr) at $G=17$ ($G=18$) corresponding to an improvement of nearly 20\% over the average precision \cite{gaia2015} of 39 $\mu$as/yr (72 $\mu$as/yr).   In our calculations below we conservatively approximate this improvement to be 15\% and so the final precision with which proper motions can be measured is
\beq
\sigma^m_p=0.85\sigma^{m0}_p. \label{propfinale}
\eeq
This function of $G$ is drawn in Fig.~\ref{sigfig} for a typical color $V-I$.  

\begin{figure} %[!tph]
\begin{center}
\includegraphics[width=2.8in,height=1.2in]{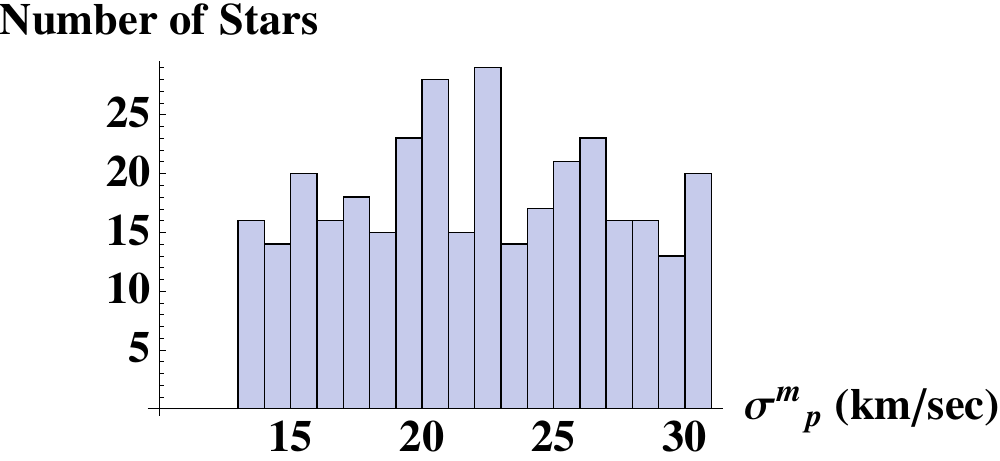}
\includegraphics[width=2.8in,height=1.2in]{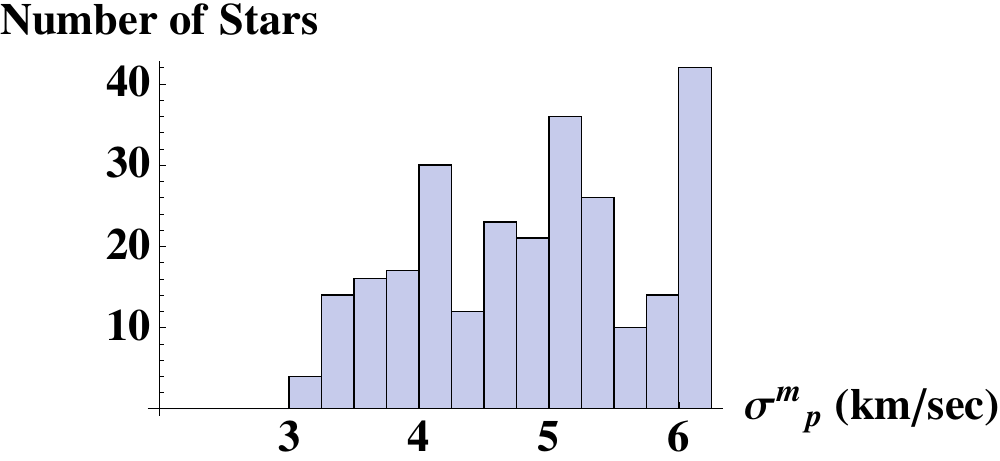}
\caption{Number of stars in Sculptor whose proper motion can be measured with a given precision by Gaia alone (top) and Gaia with TMT (bottom).  The inclusion of TMT improves the astrometric precison by about a factor of 4.}
\label{quante}
\end{center}
\end{figure}

As is explained in Ref.~\cite{posdep}, at Sculptor's ecliptic latitude of $36.5^\circ$ south this improvement is a combination of a geometric parallax factor and a higher than average number of transits.  However, the number of focal plane transits is also unusually high for Sculptor's ecliptic latitude, as a result of its longitude and Gaia's transit pattern.   Assuming a distance to Sculptor of 79 kpc \cite{mateodist}, the resulting proper motion precisions are summarized in the top panel of Fig.~\ref{quante}.

We will also determine the precision with which the proper motion can be obtained by Gaia including a survey of Sculptor by TMT in 2022 out to a radius of 700 pc using 10 second exposures of IRIS.   This will allow an astrometric precision of better than 20 $\mu$as down to magnitude $K_{AB}=20$, which includes the stars whose positions are well-measured by Gaia.  IRIS has recently be redesigned to have a $34^{\verb+"+}\times 34^{\verb+"+}$ field of view, so the required observing time will be 25 hours.  In our analysis below, we make the conservative assumption that the astrometric precision at TMT will be precisely 20 $\mu$as, whereas Gaia determines positions with a precision of $1.41\sigma^m_p$ $\mu$as \cite{gaiaperf}.  The combination of Gaia and TMT separated by 6 years therefore determines the proper motion with a precision of
\beq
\sigma^{\rm{TMT}}_p=\frac{\sqrt{400+2(\sigma^m_p)^2}\ \mu{\rm{as}}}{6{\rm{\ years}}}
\eeq
leading to the precisions summarized in the bottom panel of Fig.~\ref{quante}.  

One observes that the inclusion of TMT improves the astrometric precision by about a factor of 4, an approximation which we will adopt in our analysis.  We have verified that, as the astrometric precisions are now smaller than the dispersion, the uncertainty with which the velocity dispersion can be measured is reasonably insensitive to these precisions. 

In this letter we will answer the following questions: {\it{Given the line of sight velocities of the 1355 members in the MMFS catalog and proper motions measured with the precisions shown in Fig.~\ref{quante}: {\bf{(1)}}~To what extent can Sculptor's dark matter density profile be determined? {\bf{(2)}}~To what extent could the CDM paradigm in principle be falsified?}}

While there are many approaches to this problem in the literature, such as Refs.~\cite{penpops,agnello,pianob}, our approach to these questions will be similar to that of Ref.~\cite{strigarifisher}.  However, while we still use an assumed stellar profile to determine the expected dispersion, we perform our fit using stars in the MMFS catalog with, as described above, proper motions determined with a precision given by Eq.~(\ref{propfinale}).  The dark matter profile $\rho(r)$ is assumed to lie in a 5-parameter family, corresponding to the generalization \cite{zhaomod} of the Hernquist profile \cite{hernquistmod}
\beq
\rho(r)=\rho_0 \left(\frac{r_0}{r}\right)^a\left(1+\left(\frac{r}{r_0}\right)^b\right)^{(a-c)/b} \label{modello}
\eeq
where $r$ is the distance from the center of Sculptor.  

We assume both a constant orbital anisotropy $\beta=1-\langle v^2_\theta\rangle/\langle v^2_r\rangle$ and a spherical halo.  Ref.~\cite{strigarifisher} finds that the effect of the first assumption is minimal.   In Ref.~\cite{nonsfer} the authors used the axisymmetric Jeans equation to determine the elipticity of 6 nearby dSphs, including Sculptor.  They found that not only are these systems not spherically symmetric, but in fact their ellipticities are far higher than those found in CDM simulations.  However, this study assumed that the radial and major-axis velocity anisotropies are equal, which in the spherical case is equivalent to assuming a vanishing anisotropy $\beta$.  Indeed, the effect of ellipticity on the line of sight velocity dispersion, as described by the authors, is quite similar to that obtained by varying $\beta$.  Thus it seems likely that the statistical significance of these high ellipticities would be greatly diminished were the restriction on the anisotropies relaxed.  This effect will be compounded by relaxing their other assumptions, such as the parallel orientations of the dark matter and stellar distributions.

Nonetheless, the results of Ref.~\cite{nonsfer} demonstrate that nonspherical CDM halos do provide a systematic modification of the dispersion profiles.  This shifts the halo parameters obtained using the spherical Jeans equations.  Our goal is not to obtain the parameters themselves, but merely to determine the precision with which they can be determined.  By assuming that the halos are spherical, we will obtain only a lower bound on the uncertainty on the halo parameters.  The true uncertainty will be increased by the degeneracy between the spherical halo parameters and the anisotropy parameters.  If only line of sight velocities are available, this degeneracy is a serious problem, invalidating such an analysis.  In a future paper, we will show that with sufficiently precise measurements this degeneracy can be broken using, in particular, the azimuthal dependence of all of the velocity dispersions.  However, given Gaia's expected precision we expect that this degeneracy will nonetheless increase the uncertainties in the halo parameters somewhat beyond those reported below.

We obtain the radial-dependence of the radial velocity dispersion $\sigma^2=\langle v^2(r)\rangle$ by integrating the Jeans equation
\beq
\sigma^2(r)=\frac{G}{\rho^s(r) r^{2\beta}}\int_{r}^{r_t} \rho^s(R)M(R)R^{2\beta-2}dR \label{sigma}
\eeq
where $M(R)$ is the integrated mass within the radius $R$, approximated to be just the dark matter mass.   $\rho^s(R)$ is the 3d stellar density at  a radius $R$, not just the density of the stars in the catalog, which is taken to follow a King profile \cite{kingprof} with core radius $r_k=0.28$ kpc and tidal radius $r_t=1.63$ kpc \cite{strig2006}.  The observed line of sight, radial and tangential dispersions can be found by integrating (\ref{sigma}) over the line of sight
\bea
\sigma_{los}^2(r)\hspace{-0.2cm}&=&\hspace{-0.2cm}\frac{2}{\rho^s_{2D}(r)}\int_r^{\infty}\hspace{-0.2cm} \left(1-\beta \frac{r^2}{R^2}\right)\frac{R\rho^s(R)\sigma^2(R)dR}{\sqrt{R^2-r^2}}\nonumber\\ 
\sigma_{r}^2(r)\hspace{-0.2cm}&=&\hspace{-0.2cm}\frac{2}{\rho^s_{2D}(r)}\int_r^{\infty}\hspace{-0.2cm} \left(1\hspace{-0.1cm}-\hspace{-0.1cm}\beta\hspace{-0.1cm}+\hspace{-0.1cm}\beta \frac{r^2}{R^2}\right)\frac{R\rho^s(R)\sigma^2(R)dR}{\sqrt{R^2-r^2}}\nonumber\\ 
\sigma_{t}^2(r)\hspace{-0.2cm}&=&\hspace{-0.2cm}\frac{2}{\rho^s_{2D}(r)}\int_r^{\infty} \left(1-\beta\right)\frac{R\rho^s(R)\sigma^2(R)dR}{\sqrt{R^2-r^2}} 
\eea
where $\rho^2_{2D}(r)$ is the stellar density integrated along the line of sight at $r$, the distance from the center of Sculptor in the transverse plane.

To determine the precision with which the mass profile can be determined, we use the Fisher matrix
\beq
F_{ab}=\frac{1}{2}\sum_{i,n}\left(\frac{1}{\left(\sigma_{n}^2(r_i)+\sigma_{i,n}^{m2}\right)^2}\frac{\partial\sigma^2_{n}(r_i)}{\partial \theta_a}\frac{\partial\sigma^2_{n}(r_i)}{\partial \theta_b}\right)
\eeq
where the index $i$ runs over the stars in the MMFS catalog, the index $n$ runs over the 3 directions $los,\ r$ and $t$, $r_i$ is the projected distance from the center of Sculptor to the $i$th star, $\theta_a$ are the 5 parameters of the dark matter mass model (\ref{modello}) and $\sigma^m_{i,n}$ is the measurement uncertainty on the $n$th component of the velocity of the $i$th star.  The line of sight measurement uncertainties are taken to be 1 km/sec.  The derivatives with respect to the parameters $\theta_a$ are evaluated at the fiducial NFW model \cite{strig2006}
\beq
a=b=1\hsp c=3\hsp r_0=0.5\hsp \rho_0=8\hsp \beta=0
\eeq
where $r_0$ and $\rho_0$ are measured in units of kpc and $10^7\ M_\odot$/kpc${}^3$ respectively.

The parameter $\theta_a$ can be measured with a precision $\sqrt{(F^{-1})_{aa}}$.  By the chain rule, a quantity $q$ which depends upon the $\theta_a$ may be measured with a precision
\beq
\delta q=\sqrt{(F^{-1})_{ab}\frac{\partial q}{\partial\theta_a}\frac{\partial q}{\partial\theta_b}}.
\eeq
We will consider two quantities $q$.  The first is $M(r_{1/2})$, the mass within the deprojected half-light radius $r_{1/2}=375$ pc.  Using the Jeans equation, this can be determined directly from the stellar density profile and radial velocity dispersion at $r=r_{1/2}$, their derivatives, and $\beta$ and so is quite robust to different choices of dark matter density profile \cite{walkerhalf,wolf}.  

The second is
\beq
\gamma(r)=-3+4\pi r^3\frac{\rho(r)}{M(r)}
\eeq
which is essentially a weighted average of the exponents of the $r$-dependence of the density at radii between $0$ and $r$.  Note that, in the case of a power law $\rho(r)$, it yields the logarithmic slope.  In general it depends on the derivative of $M(r)$, which is determined less precisely than $M(r)$ itself, but unlike standard definitions it does not directly depend upon the second derivative of $M(r)$ and so is nonetheless fairly well constrained.  Below we will report the fractional uncertainty in $M(r_{1/2})$ and the total uncertainty in $\gamma(r_{1/2})$.

These quantities will be determined in 6 cases.  The first case just uses the line of sight velocities.   In the second case we impose the condition $c=3$ by adding a large number to $F_{cc}$.  This condition is satisfied by both cold and warm dark matter models, and in fact by any particulate dark matter model with no nongravitational long range interactions.  Thus, while it is certainly less general, if one is interested in testing CDM then it suffices to impose $c=3$ and test to see if $\gamma\sim -1$ as is suggested by dark matter simulations with the low baryonic content of the Sculptor dwarf.  

In the next two cases we include proper motion measurements from Gaia's 5 year mission, as has been described above, with and without the condition $c=3$.  In the last two cases we add a survey of Sculptor by the TMT in 2022.  %We make the rough approximation that this precision is a factor of four better than that of Gaia alone.  As can be seen in Fig.~\ref{quante} this approximation is reasonable, and we have verified that, since this precision is much smaller than the stellar dispersion, the approximation has little effect on our results.%Dividing the precision of Gaia's true position measurement (plus that of TMT), available in the Science Performance document \cite{gaiaperf}, by the time interval between 2026 and 2016, the middle of the Gaia mission, one obtains the expected precision of proper motion measurements of Sculptor's members which are observed by both Gaia and TMT.  This precision is about a factor of four better than that of Gaia alone.  %Note that without Gaia, a single TMT survey of Sculptor would only determine the positions of these stars and not their proper motions.  Such a precise proper motion measurement with TMT alone would take roughly five more years, although then TMT will also be able to measure the velocities of even fainter stars.  

These results are all summarized in Table~\ref{restab}.  We have also tried including estimates of $r_0$ and $\rho_0$ from CDM simulations, as summarized in Ref.~\cite{strig2006}.  With these constraints we find that Gaia can determine $\gamma(r_{1/2})$ with a precision of 0.57. % While these greatly improve the precision if $c=3$ is not imposed, simultaneously imposing both $c=3$ and these CDM simulation results leads to little improvement in precision beyond imposing $c=3$ alone.

\begin{table}
\begin{tabular}{l|c|c|}
&$\delta M(r_{1/2})/M(r_{1/2})$&$\delta \gamma(r_{1/2})$\\
\hline
\hline
LOS only&69\%&4.1\\
\hline
LOS only, $c=3$&18\%&1.6\\
\hline
Gaia&16\%&0.68\\
\hline
Gaia, $c=3$&10\%&0.66\\
\hline
Gaia+TMT&10\%&0.27\\
\hline
Gaia+TMT, $c=3$&7\%&0.27\\
\hline
\end{tabular}
\caption{The precision of measurements of the mass within $r_{1/2}=375$\ pc and the dark matter density slope within $r_{1/2}$ that can be expected with only LOS data and also at Gaia with and without TMT.  Precisions are given using a single component King profile for the stellar distribution with and without imposing that $c=3$.  The linearized Fisher matrix approach cannot be trusted for entries of order or greater than unity.}
 \label{restab}
\end{table}

In the standard $\Lambda$CDM paradigm one expects \cite{governato,gnatura} that a dark-matter dominated galaxy with the luminosity of Sculptor will have $\gamma$ equal to about -1 whereas other dark matter models, such as a Bose-Einstein condensate or giant monopole model, suggest $\gamma$ closer to zero.   If $\gamma$ indeed is close to zero, then by assuming the CDM condition $c=3$ and measuring $\gamma\sim 0$ one may hope to exclude CDM with 2$\sigma$ of confidence when the Gaia mission is completed in 5 years and with 4$\sigma$ of confidence when TMT begins observing in a decade.  One should note however that the CDM simulations on which these conclusions rest consider field galaxies, of which Sculptor is not an example.

In Ref.~\cite{sculpops} it was observed that the Sculptor dwarf contains at least two ancient populations of stars, a metal rich population near the center and a more spatially extended metal poor population.  %The determination of the shape of Sculptor's dark matter profile may be further improved by separately considering the dynamics of these subpopulations, each of which is in equilibrium.  
In Ref.~\cite{penpops} the authors found that the stellar density and metallicity distribution is well fit by the sum of two Plummer profiles of projected half-light radii 167 pc and 302 pc, having 53\% and 47\% of the members respectively.  We have redone our analysis using this bimodal stellar profile and have found a notable improvement in the precision with which $\gamma$ can be measured, as is summarized in Table~\ref{duetab}. 

%In Ref.~\cite{penpops} the exponent of the dark matter density was found by comparing the masses within two radii, but we have found that our quantity $\gamma$ can be determined more precisely than such a ratio when proper motion data is included.  However, it is important to note that our analysis does not use the fact that the stars in each Plummer profile are in fact separate populations, with distinct metallicities, each in equilibrium with itself.  The bimodal structure of the stellar population is only used in the calculation of the line of sight integrals of the luminosities and dispersions.  It would be interesting in the future to perform a true multi-population analysis.

\begin{table}
\begin{tabular}{l|c|c|}
&$\delta M(r_{1/2})/M(r_{1/2})$&$\delta \gamma(r_{1/2})$\\
\hline
\hline
LOS only&30\%&1.3\\
\hline
LOS only, $c=3$&25\%&0.86\\
\hline
Gaia&13\%&0.47\\
\hline
Gaia, $c=3$&13\%&0.44\\
\hline
Gaia+TMT&9\%&0.22\\
\hline
Gaia+TMT, $c=3$&6\%&0.21\\
\hline
\end{tabular}
\caption{As in Fig.~\ref{restab} but using a 2-component Plummer model for the stellar density profile.}
\label{duetab}
\end{table}

If one trusts the equilibrium approximation even for fourth moments of the velocities, then these may be included without introducing any new free parameters following the strategy of Ref.~\cite{malquarto}.  Alternately, as this strategy introduces more observables without increasing the number of unknowns, the proper motion measurements may be used to test the consistency of the fourth order equilibrium analysis, which may in turn shed light on Sculptor's formation.

Observation of a core in Sculptor cannot rule out WIMPs nor place interesting bounds on their interactions, this would require the observation of a core in an ultra faint dwarf (UFD).  Gaia cannot precisely observe sufficiently faint stars to help TMT with this goal.  However, in Ref.~\cite{hubble} the Hubble Astrometry Initiative has been proposed in which Hubble would precisely measure the locations of stars in UFDs so that future observatories, such as TMT, may determine their proper motions.  This suggests that deep observations of the Bootes I UFD by Hubble followed by observations by TMT will provide a powerful test of the WIMP paradigm.

\section* {Acknowledgement}

\noindent
I am very grateful to Manoj Kaplinghat for encouragement and guidance throughout this project and to Malcolm Fairbairn for very useful discussions.  I am supported by NSFC MianShang grant 11375201.   
%%%%%%%%%%%%%%%%%%%%%%%%%%%%%%%%


\begin{thebibliography}{99}%\setlength{\itemsep}{-2.3mm}

%%%%%%%%%%%%%%%%%%%%%%%%%%%%%%%%%

\bibitem{governato}
  F.~Governato, A.~Zolotov, A.~Pontzen, C.~Christensen, S.~H.~Oh, A.~M.~Brooks, T.~Quinn and S.~Shen {\it et al.},
  ``Cuspy No More: How Outflows Affect the Central Dark Matter and Baryon Distribution in Lambda CDM Galaxies,''
  Mon.\ Not.\ Roy.\ Astron.\ Soc.\  {\bf 422} (2012) 1231
  [arXiv:1202.0554 [astro-ph.CO]].

\bibitem{gnatura}
  A.~Pontzen and F.~Governato,
  ``Cold dark matter heats up,''
  Nature {\bf 506} 171
  [arXiv:1402.1764 [astro-ph.CO]].

\bibitem{pencons}
 J.~Penarrubia, A.~Pontzen, M.~G.~Walker and S.~E.~Koposov,
  ``The coupling between the core/cusp and missing satellite problems,''
  Astrophys.\ J.\  {\bf 759} (2012) L42
  [arXiv:1207.2772 [astro-ph.GA]].

\bibitem{strigarifisher}
 L.~E.~Strigari, J.~S.~Bullock and M.~Kaplinghat,
  ``Determining the Nature of Dark Matter with Astrometry,''
  Astrophys.\ J.\  {\bf 657} (2007) L1
  [astro-ph/0701581].


\bibitem{mmfs}
 M.~G.~Walker, M.~Mateo and E.~Olszewski,
  ``Stellar Velocities in the Carina, Fornax, Sculptor and Sextans dSph Galaxies: Data from the Magellan/MMFS Survey,''
  Astron.\ J.\  {\bf 137} (2009) 3100
  [arXiv:0811.0118 [astro-ph]]. 

\bibitem{gmag}
  C.~Jordi, M.~Gebran, J.~M.~Carrasco, J.~de Bruijne, H.~Voss, C.~Fabricius, J.~Knude and A.~Vallenari {\it et al.},
  ``Gaia broad band photometry,''
  Astron.\ Astrophys.\  {\bf 523} (2010) A48
  [arXiv:1008.0815 [astro-ph.IM]].

\bibitem{gaiaperf}
The Gaia ``Science Performance" document is available on the mission website http://www.cosmos.esa.int/web/gaia/ .

\bibitem{gaia2015}
J. H. J. de Bruijne, K. L. J. Rygl and T.~Antoja,  
``Gaia Astrometric Science Performance - Post-Launch Predictions," 
arXiv:1502.00791 [astro-ph.IM]

\bibitem{posdep}
J.~de Bruijne, 
``Astrometry with Gaia: what can be expected," talk given at the Commission 8 Science Meeting, IAU, Beijing on Aug 29, 2012.

\bibitem{mateodist}
  M.~Mateo,
  ``Dwarf galaxies of the Local Group,''
  Ann.\ Rev.\ Astron.\ Astrophys.\  {\bf 36} (1998) 435
  [astro-ph/9810070].


\bibitem{penpops}
 M.~G.~Walker and J.~Penarrubia,
  ``A Method for Measuring (Slopes of) the Mass Profiles of Dwarf Spheroidal Galaxies,''
  Astrophys.\ J.\  {\bf 742} (2011) 20
  [arXiv:1108.2404 [astro-ph.CO]].

\bibitem{agnello}
  A.~Agnello and N.~W.~Evans,
  ``A Virial Core in the Sculptor Dwarf Spheroidal Galaxy,''
  Astrophys.\ J.\  {\bf 754} (2012) L39
[arXiv:1205.6673 [astro-ph.GA]].

\bibitem{pianob}
 T.~Richardson, D.~Spolyar and M.~Lehnert,
  ``Plan beta: Core or Cusp?,''
  Mon.\ Not.\ Roy.\ Astron.\ Soc.\  {\bf 440} (2014) 1680
  [arXiv:1311.1522 [astro-ph.GA]].

\bibitem{zhaomod}
  H.~Zhao,
  ``Analytical models for galactic nuclei,''
  Mon.\ Not.\ Roy.\ Astron.\ Soc.\  {\bf 278} (1996) 488
  [astro-ph/9509122].

\bibitem{hernquistmod}
  L.~Hernquist,
  ``An Analytical Model for Spherical Galaxies and Bulges,''
  Astrophys.\ J.\  {\bf 356} (1990) 359.
  
\bibitem{nonsfer}
  K.~Hayashi and M.~Chiba,
  ``Probing non-spherical dark halos in the Galactic dwarf galaxies,''
  Astrophys.\ J.\  {\bf 755} (2012) 145
  [arXiv:1206.3888 [astro-ph.CO]].



\bibitem{kingprof}
  I.~King,
  ``The structure of star clusters. I. An Empirical density law,''
  Astron.\ J.\  {\bf 67} (1962) 471.

\bibitem{strig2006}
  L.~E.~Strigari, S.~M.~Koushiappas, J.~S.~Bullock and M.~Kaplinghat,
  ``Precise constraints on the dark matter content of Milky Way dwarf galaxies for gamma-ray experiments,''
  Phys.\ Rev.\ D {\bf 75} (2007) 083526
  [astro-ph/0611925].

\bibitem{walkerhalf}
 M.~G.~Walker, M.~Mateo, E.~W.~Olszewski, J.~Penarrubia, N.~W.~Evans and G.~Gilmore,
  ``A Universal Mass Profile for Dwarf Spheroidal Galaxies,''
  Astrophys.\ J.\  {\bf 704} (2009) 1274
   [Erratum-ibid.\  {\bf 710} (2010) 886]
  [arXiv:0906.0341 [astro-ph.CO]].

\bibitem{wolf}
  J.~Wolf, G.~D.~Martinez, J.~S.~Bullock, M.~Kaplinghat, M.~Geha, R.~R.~Munoz, J.~D.~Simon and F.~F.~Avedo,
  ``Accurate Masses for Dispersion-supported Galaxies,''
  Mon.\ Not.\ Roy.\ Astron.\ Soc.\  {\bf 406} (2010) 1220
  [arXiv:0908.2995 [astro-ph.CO]].

\bibitem{sculpops}
E.~Tolstoy, M.~J.~Irwin, A.~Helmi, G.~Battaglia, P.~Jablonka, V.~Hill, K.~A.~Venn and M.~D.~Shetrone {\it et al.},
  ``Two distinct ancient components in the Sculptor dwarf spheroidal galaxy: First results from DART,''
  Astrophys.\ J.\  {\bf 617} (2004) L119
  [astro-ph/0411029].

\bibitem{malquarto} T.~Richardson and M.~Fairbairn,
  ``On the dark matter profile in Sculptor: Breaking the $\beta$ degeneracy with Virial shape parameters,''
 Mon.\ Not.\ Roy.\ Astron.\ Soc.\  {\bf 441} (2014) 1584
 [ arXiv:1401.6195 [astro-ph.GA]].
 
\bibitem{hubble}
  N.~Kallivayalil, A.~R.~Wetzel, J.~D.~Simon, M.~Boylan-Kolchin, A.~J.~Deason, T.~K.~Fritz, M.~Geha and S.~T.~Sohn {\it et al.},
  ``A Hubble Astrometry Initiative: Laying the Foundation for the Next-Generation Proper-Motion Survey of the Local Group,''
  arXiv:1503.01785 [astro-ph.GA].
  

\end{thebibliography}
\end{document}